\documentstyle[pra,aps,epsf]{revtex}

\begin{document}

\draft
\title{\bf Time of arrival in the presence of interactions}
\author{  J. Le\'on$^a$, J. Julve$^a$, P. Pitanga$^b$ and  F. J. de
Urr\'{\i}es$^c$}
\address{$^a$ Instituto de Matem\'aticas y F\'{\i}sica Fundamental,
CSIC, Serrano 113-bis, 28006 MADRID, Spain}
\address{$^b$ Instituto de F\'\i sica, UFRJ. 21945-970
  Rio de Janeiro - Brasil}
\address{$^c$  Departamento de F\'{\i}sica, Universidad
de Alcal\'a, Alcal\'a de Henares, Spain}
\date{\today}
\maketitle

\begin{abstract}
  We introduce a formalism for the calculation of the
time of arrival $t$ at a space point for particles traveling
through interacting media. We develop a general formulation that
employs quantum canonical transformations from the free to the
interacting cases to construct $t$ in the context of the Positive
Operator Valued Measures. We then compute the probability
distribution in the times of arrival at a point for
 particles that have undergone reflection, transmission or tunneling
off finite potential barriers. For narrow Gaussian initial wave
packets we obtain multimodal time distributions of the reflected
packets and a combination of the Hartman effect with unexpected
retardation in tunneling. We also employ explicitly our formalism
to deal with arrivals in the interaction region for the step and
linear potentials.
 \end{abstract}



\section{Introduction}
In this paper we work out a theoretical framework to compute the
time in which a particle that moves through an interacting medium
arrives at a given point.  In the construction of this framework we
will have to deal with problems of very different kind that we
introduce now:

First, there is the nature of time in quantum mechanics. It appears
as the external evolution parameter in the Schr\"odinger and
Heisenberg equations, common to both, systems and observers alike.
However, time arises in many instances (transitions, decays,
arrivals, etc.) as a property of the physical systems. The attempts
to promote time to the category of observable run early into the
obstruction detected by Pauli~\cite{Pauli}: A self-adjoint time
operator implies an unbounded energy spectrum. This was soon
related to the uncertainty relation for time and energy, whose
status and physical meaning produced some
controversies~\cite{Mandelstamm,Aharonov1,Fock,Aharonov2}, and is
still subject of elucidation today (see for instance~\cite{Busch1}
and~\cite{Hilgevoord}). The question remains unsettled for closed
quantum systems, specially in the case of quantum gravity, whose
formulation is pervaded by the so called {\sl problem of
time}~\cite{Isham}.

Second, the definition of the time-of-arrival ({\bf toa}), which is
probably the simplest candidate time to become a property of the
(arriving) physical system, rather than a mere external parameter.
Due to its conceptual simplicity, it has been used in many cases to
illustrate different problems related to the role of time in
quantum theory. Allcock analyzed~\cite{Allcock} extensively the
difficulties met by the {\bf toa} concluding that they were
insurmountable. The present situation is ambiguous. On the one
side, there are theoretical analysis~\cite{Aharonov3,Oppenheim1} of
the {\bf toa} suggesting that it can not be precisely defined and
measured in quantum mechanics. This contradicts the possibility of
devising high efficiency absorbers~\cite{Brouard},  that could be
used as almost ideal detectors for {\bf toa}~\cite{Muga5}. On the
other hand, there are explicit constructions of a self-adjoint
(albeit in a pre-Hilbert space) {\bf toa} operator for the
non-relativistic free particle in one space
dimension~\cite{Tate,Delgado}, and for the relativistic free
particle in 3-D~\cite{Juan}, both avoiding the Pauli problem. There
is also an alternative formulation~\cite{Giannitrapani1}  as a
 Positive Operator Valued Measure (POVM). Finally, the {\bf toa} has been
 measured in high precision experiments~\cite{Steinberg3,Steinberg4}
 on the arrival of two entangled photons produced by parametric
 down-conversion, one of which has undergone tunneling through a
 photonic band gap (PBG). The experimental results that show superluminal
 tunneling, neatly identify the Hartman effect~\cite{Hartman}
 and the Wigner time delay~\cite{Wigner} (or phase time) as the
 physically relevant mechanisms for the tunneling time and {\sl
 toa} respectively. Whether these results apply only to photons and
 are due to the specific properties of the PBG used, or can be extended
 to other particles and barriers, can not be decided in the lack of a
 satisfactory theory of the {\bf toa} at a space point through
 interacting media.

The third question is thus the tunneling time, for which there are
three main proposals. Wigner introduced the phase time in his
analysis~\cite{Wigner} of the relationship between retardation,
interaction range, and scattering phase shifts. Buttiker and
Landauer introduced the traversal time~\cite{Buttiker1} in their
study of tunneling through a time-dependent barrier. Soon after,
Buttiker used the Larmor precession as a clock~\cite{Buttiker2},
identifying the dwell~\cite{Smith}, traversal, and reflection times
as three characteristic times describing the interaction of
particles with a barrier. Recent reviews that include these and
other approaches, discussing {\bf toa} and tunneling times from a
modern, unified perspective, can be found in~\cite{Muga4}
and~\cite{Muga6}. The light shed on these questions by the two
photon experiments is revised in~\cite{Chiao1} and~\cite{Chiao3}.

The main progress, quoted before, towards the formulation of a
quantum {\bf toa} operator has been its explicit construction for a
particle moving freely in one space dimension. In this paper we
face the problem of extending this formalism to the case of the
presence of an interaction potential affecting a region of the
(1-D) space. We extend here our previous unpublished
results~\cite{leon2}, taking now proper care of the dependence on
the arrival position $x$ that we consider placed in front of the
interaction region, within it, and behind it (if spatially finite).

The plan of the work is as follows: In Section II we construct a
{\bf toa} formalism of general validity. The starting point is the
case of the free particle. There, a suitable canonical
transformation, the quantum version of the Jacobi-Lie
transformation of classical mechanics, gives the {\bf toa} in
interacting media, (even at points where $V(x)\neq 0$). In Section
III we consider an initial state consisting of a narrow Gaussian
wave packet prepared at the left of the interaction region and
moving towards it. A quasi-classical study of the {\bf toa} at a
point $x$ in the interaction region is first carried out. Then we
turn to the full quantum mechanical treatment. We first analyze the
arrival in the presence of a step potential. Section IV is devoted
to the study of the {\bf toa} at points behind square barriers. We
detect, in diferent instances, stauration of and departures from
the Hartman effect. The case of $x$ at the left of the interacting
region is characterized by the (possibly interfering) contributions
coming from the incident and the reflected wave packets. This
situation is treated in Section V, where we deal separately with
the case of total reflection (very high barriers) which has an
analog in classical mechanics, and with the case of partial
reflection, a pure quantum phenomenon with very rich structure in
the time domain. Finally, we summarize our results in Section VI.

In Appendix A we show how the {\bf toa} can be treated as a derived
quantity in the phase space of classical Hamiltonian systems in the
case where these are integrable. A short review of the construction
and properties of the quantum {\bf toa} operator for free particles
is presented in Appendix B for completness.

\section{Time of arrival formalism}
To measure the time of arrival of a free particle at a point $x$
one would: a) place a detector at $x$,  b) prepare the initial
state $|\psi\rangle$ of the particle at $t=0$, and then, c)  record
with a clock the time $t$ when the detector clicks. The value of
$t$ gives the {\bf toa} of the state $|\psi\rangle$ at $x$.
Repeating this procedure with identically prepared initial states,
one would get the probability distribution in times of arrival at
$x$. Of course, the results would depend on the initial state
chosen, which stores all the information regarding the initial
distribution in positions and momenta of the particle.

We want to determine the effect on these times of a position
dependent interaction between the particle and the medium, that we
describe by a potential energy $V(q)$. For instance, to disclose
the effect of climbing (or tunneling through) a potential barrier,
one would simply put the barrier in between the detector and the
initial state, and then record the new times of arrival. With an
initial state identical to that prepared for the free case, any
difference in the probability distributions should be an effect of
the barrier. Several questions can be investigated by changing the
properties of the barrier: its height or width if it is
rectangular, even its very form.  This has been explicitly done in
the two photon experiments at Berkeley, by putting alternatively a
mirror and an ordinary glass in the path of one of the photons. Of
great interest is the dependence on $x$ where $V(x)>0$, i.e. with
the detector within the range of the interaction, and also the time
of arrival for $E>V$, in which classically there is no reflection.
Some of these questions are studied in last sections of this paper.

In classical mechanics particles move along the trajectories
$H(q,p)=$const. as $t$ increases. This allows to work out   $t_x$,
the time of arrival at the point $q(t)=x$, by identifying the point
$(q,p)$ of phase space where the particle is at (say) $t=0$,  and
then by following the trajectory that passes by it, up to the arrival
at $x$. The mathematical translation of this procedure is given by
the equation of time:
\begin{equation}
t_x(q,p)=\mbox{sign}(p) \sqrt{\frac{m}{2}}\int_{q}^{x}\frac{dq'}
      {\sqrt{H(q,p)-V(q')}}\label{j5}
\end{equation}
that is discussed at length in many textbooks, and whose existence
conditions and characterization as a function of the phase space
variables are outlined in the Appendix A. We simply note here that
$t_x(q,p)$ is canonically conjugate to the hamiltonian
$\{t_x(q,p),H(q,p)\}=-1$.

This equation is a troublesome starting point for quantization.
First, it involves a (path) integral of operators and should be
treated accordingly. Second, it only applies to values of $x$ that
are classically within the reach of $(q,p)$, while in quantum
mechanics all values of $x$ are attainable. Classically, the
particle propagates without reflection up to the turning point
$q_0$ ($V(q_0)=E$), where it is completely reflected. There is no
further penetration beyond this point. The situation is different
in quantum mechanics: there may be tunneling beyond $q_0$ and
partial reflection before reaching it. These phenomena cannot be
accounted for by Eq. (\ref{j5}), that gives complex numbers for
these cases. Now, note that both, tunneling and partial reflection,
are absent from the motion of free particles, whose time of arrival
has been successfully quantized as said in the introduction. In
addition,  all the positions are within the reach of the free
particle. Summarizing, everything points to the free time of
arrival as a main clue to solve the problem.

In this work we desist from  attempting the straightforward
quantization of the classical expression (\ref{j5}). Instead, we
will construct the solution to the interacting case taking as
starting point the well known results that apply to the free case.
The aim is to produce the quantum version of the Lie transformation
from the actual flow in phase space to the canonically equivalent
parallel flow of constant velocity translations. In other words, we
shall use the quantum version of the canonical transformation to
action-angle variables. The Lie procedure -- that we sketch for
completness in Appendix A -- has a property that will be the crux
of the matter in our construction. Namely, it permits to define
time as a derived variable in phase space in terms of the free
action-angle variables as well as, alternative and equivalently, in
terms of the original positions and momenta. Obviously, both
definitions give the same result as we show explicitly in Eq.
(\ref{classic5}). Our use of the Lie procedure in the quantum case
can be described as the combination of steps a,b and c below:
\begin{itemize}
\item[a.] The quantization of the time of arrival ${\bf t_0}$ of the
free particle. This is an old problem in quantum mechanics, whose
solution in terms of a Positive Operator Valued Measure we describe
in Appendix B.
\item[b.] The construction of the quantum canonical transformation
$U$ that
connects the free particle dynamics with Hamiltonian $H_0$ to the
case of interest with Hamiltonian $H_0+V(q)$. $U$ is given by the
M\"oller wave operator as we show in Sections 2.1 and 2.2.
\item[c.] The application of the canonical transformation $U$
to ${\bf t_0}$
to get the time of arrival ${\bf t}$ in the presence of the
interaction potential $V(q)$, that is ${\bf t}=U\,{\bf t_0}\,
U^\dagger$. This is what we do in Section 2.2, where we also
address the interpretation of the resulting formalism.
\end{itemize}

\subsection{Implicit quantum canonical transformations}
Classical canonical transformations $\bar q=\bar q (q,p)$, $\bar
p=\bar p(q,p)$ in phase space can be defined implicitly by the use
of auxiliary functions $F,G,\bar F,\bar G$ in the following way:
\begin{eqnarray}
\bar F(\bar q,\bar p)& =&F(q,p)\nonumber\\ \bar G(\bar q,\bar p)&
=&G(q,p) \label{k0}.
\end{eqnarray}
It is easy to work out the following relation among Poisson
brackets:
\begin{equation}
\{\bar F,\bar G\}_{\bar q \ \bar p}  \{\bar q,\bar p\}_{q \
p}=\{F,G\}_{q\ p} \label{k1}
\end{equation}
In these conditions, the transformation is canonical (i.e.  $\{\bar
q,\bar p\}_{q \ p}=1$) if and only if
\begin{equation}
\{\bar F,\bar G\}_{\bar q \ \bar p}=\{F,G\}_{q\ p} \label{k2}.
\end{equation}
 This relation has the additional property of fixing one of the four
 functions $F,G,\bar F,\bar G$, once the other three are given.
 We can choose $ F$ and $ G$ as the free particle Hamiltonian and time
 of arrival respectively. Then, if $\bar F$ is the complete Hamiltonian
 $H$, $\bar G$ will be the corresponding {\bf toa} $t_x$ given by
 (\ref{j5}) along the classical trajectories.

Canonical transformations were introduced by Dirac in quantum
mechanics~\cite{Dirac} by the use of unitary transformations $U$
($U U^\dagger =U^\dagger U=1$ ). If the operators $\bar {\bf q},
\bar {\bf p}$ are canonically transformed from $ {\bf q},  {\bf p}$
, then there is a unitary transformation $U$ such that
\begin{eqnarray}
\bar {\bf q}&=&U^\dagger {\bf q} U \nonumber\\ \bar {\bf
p}&=&U^\dagger {\bf p} U. \label{k4}
\end{eqnarray}
Then one can define implicitly quantum canonical transformations,
like the classical ones. This possibility has been thoroughly
analyzed and  developed. The main results of the method are
collected in~\cite{Moshinsky3}, where one can also find references
to other relevant literature. The transformation $U$ is given by
\begin{eqnarray}
\bar {\bf F}(\bar {\bf q},\bar {\bf p})& =&U^\dagger\bar {\bf
F}({\bf q},{\bf p})U ={\bf F}({\bf q},{\bf p}) \nonumber \\ \bar
{\bf G}(\bar {\bf q},\bar {\bf p})& =&U^\dagger\bar {\bf G}({\bf
q},{\bf p})U ={\bf G}({\bf q},{\bf p}) \label{k5}
\end{eqnarray}
where the last equality in each row is the definition of the barred
operators in terms of the unbarred ones, while the first equality
comes from the straight application of (\ref{k4}) to the l.h.s.
Being $U$ a unitary transformation, the spectra of the canonically
transformed operators have to coincide, that is:
\begin{eqnarray}
 \sigma (\bar {\bf q}) = \sigma ({\bf q}) = {\cal R},&
 \sigma (\bar {\bf p})= \sigma ({\bf p}) ={\cal R}\nonumber \\
 \sigma (\bar {\bf F})=\sigma ({\bf F}) ,& \sigma (\bar {\bf G})=
 \sigma ({\bf G})\label{k6}
\end{eqnarray}
where the second row stands because $\bf F$ and $\bar {\bf F}$,
$\bf G$ and $\bar {\bf G}$ are also unitarily related operators.

The above relations permit to build the operator $\bar {\bf G}$
once ${\bf F},{\bf G}$ and $\bar {\bf F}$ are given. We assume that
${\bf F}$ and $\bar {\bf F}$ are self-adjoint operators, with the
eigenstates corresponding to the same eigenvalue $\lambda_f$ given
by:
\begin{equation}
\bar {\bf F}|{\bar f}\rangle = \lambda_f|{\bar f}\rangle,\,\, {\bf
F}|f \rangle = \lambda_f|f\rangle \label{k7}
\end{equation}
They form orthogonal and complete bases satisfying
\begin{eqnarray}
\langle f s|f's'\rangle=&\delta_{ss'}
\delta(\lambda_f-\lambda_f'),\, \;\sum_s \int_{\sigma (\lambda)}
d\lambda_f |f s\rangle \langle f s| &={\rm I\!I}\\ \langle \bar f
s|\bar f' s'\rangle=&\delta_{ss'} \delta(\lambda_f-\lambda_f'),\,\;
\sum_s \int_{\sigma (\lambda)} d\lambda_f |\bar f s\rangle
\langle\bar f s| &={\rm I\!I}\label{k8}
\end{eqnarray}
where we allow for some degeneracy (that has to be the same for
both ${\bf F}$ and $\bar {\bf F}$) labeled by $s$. We have also
assumed that $\lambda$ is continuous, while $s$ is a discrete
index. These assumptions could be changed straightforwardly if it
were necessary. Now, an operator $U$ satisfying the first row of
Eq.(\ref{k5}) can be given simply as:
\begin{equation}
U=\sum_s \int_{\sigma (\lambda)}  d\lambda_f |\bar f s\rangle
\langle  f s| \label{knew}
\end{equation}
It is straightforward to verify that it is unitary. We can now
proceed to the sought for result: the definition of $\bar {\bf G}$
in terms of ${\bf G}$ using $U$, that is $\bar {\bf G}=U {\bf G}
U^\dagger$. The full fledged expression is
\begin{equation}
\bar {\bf G}({\bf q},{\bf p})=\sum_{ss'} \int_{\sigma({\lambda})}
d\lambda_f d\lambda_{f'}|\bar f s\rangle \, \langle f s|{\bf
G}({\bf q},{\bf p})|f's'\rangle \, \langle \bar f 's'| \label{k9}
\end{equation}
that constitutes our main result in the quantum canonical
formalism.
\subsection{Definition of the time of arrival}
We will now apply the above to the case where $\bf F$ is the free
Hamiltonian $H_0$, $\bar {\bf F}$ the complete Hamiltonian $H$ and
$\bf G$ the time of arrival of the free particle Eq.(\ref{jj}).
Then, we have $H_0=U^\dagger H U$ and $ \Pi_0(x)=U^\dagger \Pi(x)
U$. In Appendix B we have summarized the {\bf toa} formalism for
the free particle, given by the positive operator valued measure
$P_0$ of Eq.(\ref{j6}). Accordingly, the POVM $P$ of the
interacting case will be given by ({\it cf} (\ref{k5}))
\begin{equation}
P(\Pi(x);t_1,t_2 )=U P_0 (\Pi_0(x);t_1,t_2) U^\dagger. \label{k9b}
\end{equation}
Finally, the time of arrival operator in the presence of
interactions (the $\bar{\bf G}$ of our problem) is given by
\begin{equation}
{\bf t}(H,\Pi (x))=U {\bf t}_0 (H_0,\Pi_0(x)) U^\dagger.
\label{k9a}
\end{equation}

Three comments are in order here:
\begin{itemize}
\item Fixing $U$ by the relation between both hamiltonians leads
to two different solutions:
\begin{equation}
U_{(\pm)}=\sum_s \int_0^{\infty} dE |E s (\pm)\rangle \langle E s
0|=\Omega_{(\pm)} \label{k10}
\end{equation}
which are the M\"oller operators connecting the Hilbert space
 ${\cal H}_{in}$ and ${\cal H}_{out}$ of free particle states
to the Hilbert space ${\cal H}$ of the bound and scattering states.
 These operators
are only isometric in the presence of bound states, because the
correspondence between states in $\cal H$ and free states can not
be one to one. In this paper  we will consider only well behaved
potentials ($V(q)\geq 0\, \forall \, q$), that vanish at the
spatial infinity, for which the M\"oller operators are unitary
because there is one free state for each scattering state. In this
case, the intertwining relations $H \Omega_\pm=\Omega_\pm H_0$ can
be put in the usual form $H =\Omega_\pm H_0\Omega^\dagger_\pm$. In
addition, we shall adhere to the standard conventions, choosing
$\Omega_{(+)}$ (with $E=\lim_{\epsilon\rightarrow
0^+}(E+i\epsilon)$) in (\ref{k10}) that gives signal propagation
forward in time. The results that would be obtained with
$\Omega_{(-)}$ would correspond to the time reversal of the actual
situation. If $\tau$ is the time reversal operator
$P_{(-)}(\Pi(x);t_1,t_2 )=\tau \; P_{(+)}(\Pi(x);-t_2,-t_1 )\;
\tau^\dagger$. For notational simplicity, we will omit this label
$(+)$ wherever possible.

\item The reduction of the problem to a sort of free particle
problem by means of a canonical transformation as done in
(\ref{k9a}) should not be a surprise. On the contrary, this is the
quantum counterpart of the classical situation where the
trajectories of completely integrable phase space flows  can be
straightened out to those of a free particle by means of a
canonical transformation. To the classical Lie transformation of
Appendix A that carries out this stretching corresponds the quantum
transformation described above and in the previous Section 2.1.
Concretely, Equation (\ref{classic5}) is the classic analog to
(\ref{k9a}).
\item $x$ is the actual detector position in the interacting case.
Therefore, the arguments of $\bf t$ in (\ref{k9a}) have to be
$\Pi(x)= |x\rangle \langle x|$ and $H$. This gives for the argument
of ${\bf t}_0$ an object $\Pi_0(x)=\Omega^\dagger \Pi(x) \Omega$
which is not a position projector. Instead, it collects all the
states of the free particle that add up to produce the position
eigenstate $|x\rangle$ of the interacting case by the canonical
transformation. Much of the difference between the classic and
quantum cases is hidden here, in particular the quantum capability
to undergo classically forbidden jumps in phase space.
\end{itemize}

Summarizing, in the interacting case we have a {\bf toa} operator
given by
\begin{equation}
{\bf t}_x=\sum_s \int_{-\infty}^{+\infty} dt\, t \, |t x s\rangle
\langle t x s| \label{k20},
\end{equation}
where
\begin{equation}
|t x s\rangle=(\frac{2H}{m})^{1/4}\, e^{i H t}\,
 \Pi_s\, |x\rangle \label{k21}
\end{equation}
Above we have introduced the projector $\Pi_s=\int dE\, |E s
(+)\rangle \langle E s (+)|$, which is obtained from the $\Pi_{s0}$
of Eq. (\ref{k15b}) by the canonical transformation (\ref{k10}). We
now have the tools necessary for a physical interpretation in terms
of a POVM: Given an arbitrary state $\psi$ at $t=0$, its time of
arrival at a position $x$ has to be, according to (\ref{k20}),
 \begin{equation}
\langle\psi|{\bf t}_x|\psi\rangle=\frac{1}{P(x)}\sum_s
\int_{-\infty}^{+\infty} dt\, t \, |\langle t x s|
\psi\rangle|^2,\label{k22}
\end{equation}
with the standard interpretation of $\sum_s|\langle t x s|
\psi\rangle|^2$ as the (yet unnormalized) probability density that
the state $|\psi\rangle$ arrives at $x$ in the time $t$. The
probability of arriving at $x$ at any time is then $P(x)=\int dt\;
\sum_s|\langle t x s| \psi\rangle|^2$, giving a normalized
probability density in times of arrival
\begin{equation}
P(t,x)=\frac{1}{P(x)}\;\sum_s|\langle t x s|
\psi\rangle|^2\label{k222}
\end{equation}
normalization that has been used in (\ref{k22}). Note that in the
cases where $P(x)$ vanishes this conditional probability is devoid
of meaning: If there are no arrivals at all, there are no arrivals
in any finite (or infinitesimal) interval of time.

The above equations (\ref{k22},\ref{k222}) can be given a form that
is very useful for computation, while throws some light on the
physical meaning of the different quantities involved. By using
explicitly (\ref{k21}), one gets
\begin{eqnarray}
P(x)&=&\sum_s \{\int dE\,(\frac{2E}{m})^{1/4} \langle x|E s
(+)\rangle \langle E s (+)|\psi\rangle \}^*\times\nonumber\\ &&
 \{\int dE'\, (\frac{2E'}{m})^{1/4} \langle x|E' s (+)\rangle
\langle E' s (+)|\psi\rangle \} \int dt \, e^{-i (E-E')t}\nonumber
\\ &=&2 \pi \sum_s \int dE (\frac{2 E}{m})^{1/2} |\langle x|E s
(+)\rangle \langle E s (+)|\psi\rangle|^2\label{k23}
\end{eqnarray}
Using a similar procedure, one gets for (\ref{k22})
\begin{eqnarray}
\langle \psi|{\bf t}_{x}|\psi\rangle&=&-\frac{i \pi}{P(x)} \sum_s
\int dE (\frac{2 E}{m})^{1/2}\times\nonumber \\ && \{\langle x|E s
(+)\rangle \langle E s (+)|\psi\rangle \}^*\,
\stackrel{\longleftrightarrow}{\frac{d}{dE}}\,\{\langle x|E s
(+)\rangle \langle E s (+)|\psi\rangle \}\nonumber\\ &=& \frac{2
\pi}{P(x)} \sum_s \int dE (\frac{2 E}{m})^{1/2}\nonumber  |\langle
x|E s (+)\rangle \langle E s (+)|\psi\rangle |^2\\\ &&
\frac{d}{dE}\{\arg\langle x|E s (+)\rangle +\arg \langle E s
(+)|\psi\rangle\}\label{k24}
\end{eqnarray}

\section{The entrance into the interaction region}
We start here to analyze the theoretical predictions of our
formalism. To begin with, we consider the simple case of an initial
Gaussian state prepared at $t=0$ in a zone where $V(q)=0$, and
directed towards the interaction region. This wave packet $\psi$ of
 width $\Delta q=2 \delta$, is centered at $q_0<0$ -well to the left
of the onset of the interaction- with mean momentum $p_0>0$. In
configuration and momentum spaces we have:
\begin{eqnarray}
\langle q|\psi\rangle& =&(\frac{1}{2 \pi \delta^2})^{1/4}\,
e^{-\delta^2 p_0^2}\,\, e^{-(\frac{q-q_0}{2\delta}- i \delta
p_0)^2}\label{l8}\\
 \langle p|\psi\rangle& =&(\frac{2\delta^2}{\pi
})^{1/4}\, e^{-\delta^2 (p-p_0)^2-ipq_0}\nonumber
\end{eqnarray}
respectively. For appropriate values of $q_0,p_0$ and $\delta$,
such that $p_0\delta >>1$ and $|q_0|>>\delta$, almost all the
packet is initially at the left of the origin and moving with
positive momentum towards the right. We use this simplifying
assumption (the neglect of the Gaussian's tails with $q>0,p<0$)  in
our qualitative arguments, and in the intuitive descriptions of the
processes that we will develope below. This will be indicated
explicitly in the formulas by the use of $\approx$ instead of $=$.
However, we shall work with the full expressions (\ref{l8})
wherever necessary in the calculations. For simplicity, we consider
that the potential vanishes to the left of the origin. Preparing
the state $\psi$ as said above with $\psi(q)\approx 0$ for $q>0$,
and its Fourier transform $\tilde{\psi}(p)\approx0$ for $p<0$, we
have $\langle E s (+)|\psi\rangle \approx \delta_{sr} (\frac{m}{2
E})^{1/4} \tilde{\psi}(p)$, so that
\begin{equation}
\langle t x s|\psi\rangle\approx \delta_{sr} \int dE\; e^{-iEt}
\langle x|Er(+)\rangle \tilde{\psi}(p)\label{n1}
\end{equation}
valid for the full range of values of $x$. Now, the initial state
contributes to the time of arrival (\ref{k24}) a quantity $ d/dE
\arg\langle E s (+)|\psi\rangle\approx -m q_0/p$, the same that in
the free case.

\subsection{The quasi-classical case}
We start with the simple but illustrative case where the potential
departs from 0 for positive $q$ with $V(0)=0$, and is so smooth
that the WKB method is valid. Then, for $E>V(x)$ and to lowest
order, one can neglect the exponentially small reflection that
would vanish classically, getting energy eigenstates of the form
\begin{equation}
\langle x|E r (+)\rangle\approx\theta(-x)\sqrt{\frac{m}{2\pi p}}
e^{ipx}+ \theta(x)\sqrt{\frac{m}{2\pi p(x)}} e^{i\int_0^x dq\,
p(q)}\label{n3}
\end{equation}
where $p(q)=\sqrt{2m(E-V(q))}$.  To this order and with a properly
normalized wave packet as ours, (\ref{k23}) gives
\begin{equation}
P(x)\approx \theta(-x)+\theta(x) P_+(x) ,\; P_+(x)=\int_0^\infty dp
\frac{p}{p(x)}\;|\tilde{\psi}(p)|^2\label{n4}
\end{equation}
so that $\frac{p}{p(x)}\;|\tilde{\psi}(p)|^2$ is the (unnormalized)
probability of arrival at the point $x$ with momentum $p(x)$. For
the probability in times of arrival one gets
\begin{eqnarray}
P(t,x)&\approx&\frac{\theta(-x)}{2\pi}\left| \int_0^\infty dp\,
e^{-iEt}\;\tilde{\psi}(p)\right|^2 \nonumber\\
&&+\frac{\theta(x)}{2\pi P_+(x)}\left|\int_0^\infty dE\;
\sqrt{\frac{m}{p(x)}}\;\tilde{\psi}(p)\;  e^{-i(Et-\int_0^x dq \
p(q))}\;\right|^2 \label{n5}
\end{eqnarray}
which is the same as that of free particles for $x<0$ as
corresponds to this order of approximation in which reflection is
neglected, so that there is no information about $V$ at the left of
the origin. Finally,
\begin{eqnarray}
 \langle\psi|{\bf
t}_x|\psi\rangle&\approx& \theta(-x) \int_0^\infty dp\;
|\tilde{\psi}(p)|^2 \;\frac{m}{p}\{x-q_0\}\nonumber\\
&&+\frac{\theta(x)}{P_+(x)}\int_0^\infty dp\,
\frac{p}{p(x)}\;|\tilde{\psi}(p)|^2\; \{-\frac{mq_0}{p}+m \int_0^x
\frac{ dq}{p(q)}\}\label{n6}
\end{eqnarray}
Therefore, for negative $x$ we recover the {\bf toa} of the free
particle.  What the above expression gives for $x>0$ is nothing
else than the  classical time of arrival at $x$, Eq. (\ref{j5}),
for initial conditions $(q_0,p)$  weighted by the probability of
 these conditions.

\subsection{Step potential and Hartman effect}
In general, the approximations that led to (\ref{n3}) do not hold.
For instance, reflection has to be taken into account, or $V$ is
such that the semiclassical approximation is no longer valid, etc.
In any case, the particle may eventually reach a point $q$ where
$E=V(q)$. Any further penetration beyond that point is a quantum
fenomenon worth to investigate in terms of the {\bf toa}. We
address this question by considering a step potential $V(q)=\theta
(q) V$ intercepting the path of the wave packet $\psi$. We will
then analyze the fate of the components of the wave packet with
$p>p_V=\sqrt{2m V}$ and with $p<p_V$. Classically, a particle in
the first group will arrive with momentum $p'=\sqrt{|p^2-P_V^2|}$
at the points $x>0$, while one in the second group will bounce back
at $q=0$, without penetrating to the right. In the quantum case,
one has for $x<0$ a superposition of both, reflection and
transmission, regardless of $p/p_V$, while for $x>0$ one has
\begin{eqnarray}
\langle t x s|\psi\rangle &\approx&
\frac{\delta_{sr}}{\sqrt{2\pi}}\int dE (\frac{m}{2E})^{1/4} e^{-i
Et}\times\nonumber\\ &&  \{\theta(E-V)T_> e^{ip'x}+\theta(V-E)T_<
e^{-p'x}\} \tilde{\psi}(p) \label{n7}
\end{eqnarray}
where $T_>=2p/(p+p')$  and $T_<=2p/(p+ip')$. Then,
\begin{equation}
P(x)\approx\int_{p_V}^{\infty} dp\, |T_>\,\tilde{\psi}(p)|^2+
\int_0^{p_V}dp\, e^{-2p'x}|T_<\,\tilde{\psi}(p)|^2\label{n8}
\end{equation}
is the probability of arrival at $x$, while
\begin{eqnarray}
P(t,x)&\approx&\frac{1}{2\pi P(x)}\left| \int dE
(\frac{m}{2E})^{1/4} e^{-i Et}\times\right. \nonumber\\ &&\left.
\{\theta(E-V)T_> e^{ip'x}+\theta(V-E)T_< e^{-p'x}\}
\tilde{\psi}(p)\right|^2\label{n9}
\end{eqnarray}
gives the probability distribution in {\bf toa} of the particles
that arrive at this point. Finally,
\begin{eqnarray}
\langle\psi |{\bf t}_x|\psi\rangle&\approx&\frac{1}{P(x)}\left[
\int_{p_V}^{\infty} dp\,
|T_>\,\tilde{\psi}(p)|^2\{-\frac{mq_0}{p}+\frac{mx}{p'}+\frac{m}{p}\frac{d
\arg (T_>)}{dp}\}\right. \nonumber\\
 &&\left. + \int_0^{p_V}dp\,
e^{-2p'x}|T_<\,\tilde{\psi}(p)|^2\{-\frac{mq_0}{p}+\frac{m}{p}\frac{d\arg
(T_<)}{dp}\} \right] \label{n10}
\end{eqnarray}

In the case of low potential steps $p_V<<p_0$ (c.f. Eq.
(\ref{l8})), where one can neglect the integrals over the interval
$[0,p_V]$, the probability of arrival reduces to the average of the
transmision coefficient $|T_>|^2$, which is independent of $x$ as
corresponds to a transmitted free particle. $T_>$ is real in this
case, so that $\langle\psi |{\bf t}_x|\psi\rangle$ is given by
averaging over $p$ the time spent to go from  $q_0$ to 0 at
momentum $p$ plus the time spent to go from 0 to $x$ at momentum
$p'$. The only effect of the step is the reduction of the momentum
from $p$ to $p'$.

In the opposite case where $p_V>>p_0$, only the integrals over
$[0,p_V]$ give a sizeable contribution. The probability of arrival
vanishes (exponentially) beyond the distance $\Delta x =
\frac{1}{p'}$ associated through the uncertainty principle to the
difference $\Delta E$ between the energy of the step and the energy
of the particle. One then expects to detect a relative of this
fenomenon in the time of arrival. In fact, the time spent from 0 to
$x$ is given here through $\frac{m}{p}\frac{d \arg
T_<}{dp}=\frac{m}{pp'}$, which is independent of the distance $x$,
that is replaced by $\Delta x$. This is a case of  the Hartman
effect that here arises from the change
\begin{equation}
\frac{p}{p+p'} e^{ip'x}\longrightarrow \frac{p}{p+ip'}
e^{-p'x}\label{n11}
\end{equation}
in the energy eigenstates as $p$ crosses $p_V$ from above. In
short,  the effect is a consequence of the fact that the phase is
 independent of $x$ for $p<p_V$.

 In the general case one should take into account both
 contributions to (\ref{n10}). The relative importance of the
 second contribution in the rhs would depend on $p_0-p_V$
  and will always decrease
 exponentially with increasing $x$. However, a proper analysis of
this situation  calls for a description of
 particles better that that provided by first quantization and
wave packets. We will defer this question to the next section where
we discuss tunneling, the instance where the particle may reappear
again beyond some point.

\section{Arrival at the other side}
In this section we will study the modification of the times of
arrival of quantum particles that traverse potential barriers. Our
treatment deepens on the current understanding of the tunneling and
dwell times. The literature is full of ad hoc heuristic arguments
often disconnected from the standard mathematical and
interpretative apparatus of quantum mechanics, whose value is
therefore difficult to asses, as is their comparison with
experiment. Here, we will follow the standard quantum mechanical
treatment of Section 2.

The time of arrival at a point $x$ will now be given through a
probability amplitude
\begin{equation}
\langle t x s|\psi\rangle = \int dE (\frac{2E}{m})^{1/4} e^{-i
Et}\, \langle x|Es(+)\rangle \langle Es(+)|\psi\rangle \label{psi2}
\end{equation}
We prepare the initial state as usual (as a right mover at the left
of the barrier, c.f. above Eq. (\ref{n1})). We again can
approximate $\langle E s (+)|\psi \rangle \approx \delta_{rs}
(\frac{m}{2E})^{1/4} \tilde{\psi}(p)$. The scattering state of
relevance in (\ref{psi2}) is given by
\begin{eqnarray}
\langle q | E r {(+)} \rangle&=&\sqrt{\frac{m}{2 \pi p}} \theta(-q)
(e^{i p q} + R(p) e^{-i p q})+ \theta(q) \theta(a-q)
A(q,p)\nonumber
\\&+&\theta(q-a) T(p) e^{i p q} \label{s1}
\end{eqnarray}
Expression valid for an arbitrary potential barrier contained in the
range $(0,a)$, where $A(q,p)$ solves the appropriate Schr\"odinger
equation with energy $E=p^2/2m$. Also, $T(p)$ and $R(p)$ are the
transmission and reflection coefficients of the barrier. For a
barrier of infinite range, the first and third terms in the rhs of ({\ref{s1})
should be better understood as asymptotic limits.

Finally, in the case where $x$ is at the right of the barrier, the
amplitude can be approximately given by
\begin{equation}
\langle t x s|\psi\rangle \approx
\frac{\delta_{sr}}{\sqrt{2\pi}}\int dE (\frac{m}{2E})^{1/4} e^{-i
(Et-px)} T(p) \tilde{\psi}(p)\label{psi3}
\end{equation}
The normalized probability density in times of arrival at $x$
counts all the particles eventually recorded at $x$ and only them,
that is, the transmitted particles. According to (\ref{k222}) it is
given by
\begin{eqnarray}
P(t,x)&=&\frac{1}{P(x)} \sum_s |\langle t x s|\psi\rangle|^2
\label{psi4}
\\ &\approx&\frac{1}{2\pi P(x)}\left| \int  dE (\frac{m}{2E})^{1/4} e^{-i
(Et-px)}\; T(p)\; \tilde{\psi}(p)\right|^2\nonumber
\end{eqnarray}
where we have normalized dividing by $P(x)$, the total probability
of arrival at $x$ in whatever time $t$
\begin{equation}
P(x)=\sum_s \int_{-\infty}^{+\infty} dt\;  |\langle t x
s|\psi\rangle|^2\approx \int_0^{+\infty} dp\; |T(p)\;
\tilde{\psi}(p)|^2 \label{psi44}
\end{equation}
that is independent of $x$ in cases like this, where $x$ is beyond
the range of the potential. In addition, it approximately
simplifies to $|T(p_0)|^2$ for narrow wave packets with mean
momentum $p_0$ not too close (by above or by below) to the barrier
momentum $p_V=\sqrt{2mV}$. After a straightforward calculation we
get for the average time of arrival at the other side of the
barrier
\begin{eqnarray}
\langle\psi|{\bf t}_x|\psi\rangle&\approx&- \frac{i}{2 P(x)}\times
\label{psi5}\\ && \int dE \left[(\frac{m}{2E})^{1/4} e^{ipx} T(p)
\tilde{\psi}(p)\right]^* \stackrel{\leftrightarrow}{\frac{d}{dE}}
\left[(\frac{m}{2E})^{1/4} e^{ipx} T(p)
\tilde{\psi}(p)\right]\nonumber
\end{eqnarray}
that can be written as
\begin{equation}
\langle\psi|{\bf t}_x|\psi\rangle\approx  \frac{1}{ P(x)}
\int_0^{\infty} dp\; |T(p)\, \tilde{\psi}(p)|^2 \frac{m}{p}
\{x-q_0+\frac{d\arg(T(p))}{dp}\} \label{psi55}
\end{equation}
an expression thas has appeared before in the literature sometimes
supported by heuristic arguments alone. It can be understood as the
average value of the Wigner time~\cite{Wigner} over the transmitted
state.

We will illustrate the predictions of the formalism for a simple
square barrier of height $V$ and width $a$. The transmission
coefficient is in this case:
\begin{equation}
T(p)=\frac{2 p p' e^{-i p a}}{2 p p' \cos p'a-i (p^2 +p'^2 ) \sin
p'a}\label{t1}
 \end{equation}
 where $p'=\sqrt{p^2-p_V^2}$, that is imaginary for $p$ below
 $p_V$. Note the contribution $-pa$ to the phase of $T(p)$. This
 will substract a term $a$ to the path length $x-q_0$ that
 appears in (\ref{psi55}). The barrier has effective zero width or,
 in other words, it is traversed instantaneously. This is the
 Hartman effect for barriers. To be precise, the effect is not
 complete, it is compensated by the other dependences in $p'a$ present
 in the phase of $T(p)$. In fact, it dissapears for
 $(p_V/p)\rightarrow 0$, where all the $a$ dependences of the phase
 cancel out, as was to be expected because the barrier effectively
 vanishes in this limit. In the opposite case $(p/p_V)\rightarrow 0$
 the effect saturates and there is an advance $ \frac{m a}{p}$ in the
time of arrival of transmitted plane waves, that turns into
unexpected results for intermediate barrier momenta.

 We present our results for the time of arrival of the transmitted
 particles in Figs. 1 and 2. We consider the same initial state  in
 both cases, namely the Gaussian wave packet of (\ref{l8}) with
 $q_0=-30,p_0=2,\delta=10$, and $m=1$, (we always use the natural
 units of the problem with $\hbar=1$). We have computed the time of
 arrival of the wave packet at $x=50$ for an assortment of
 potential heights and widths, and have chosen the contents of
 those figures to highlight the most important results.
 \begin{figure} \epsfxsize=16cm
 \begin{center}
 \ \epsffile{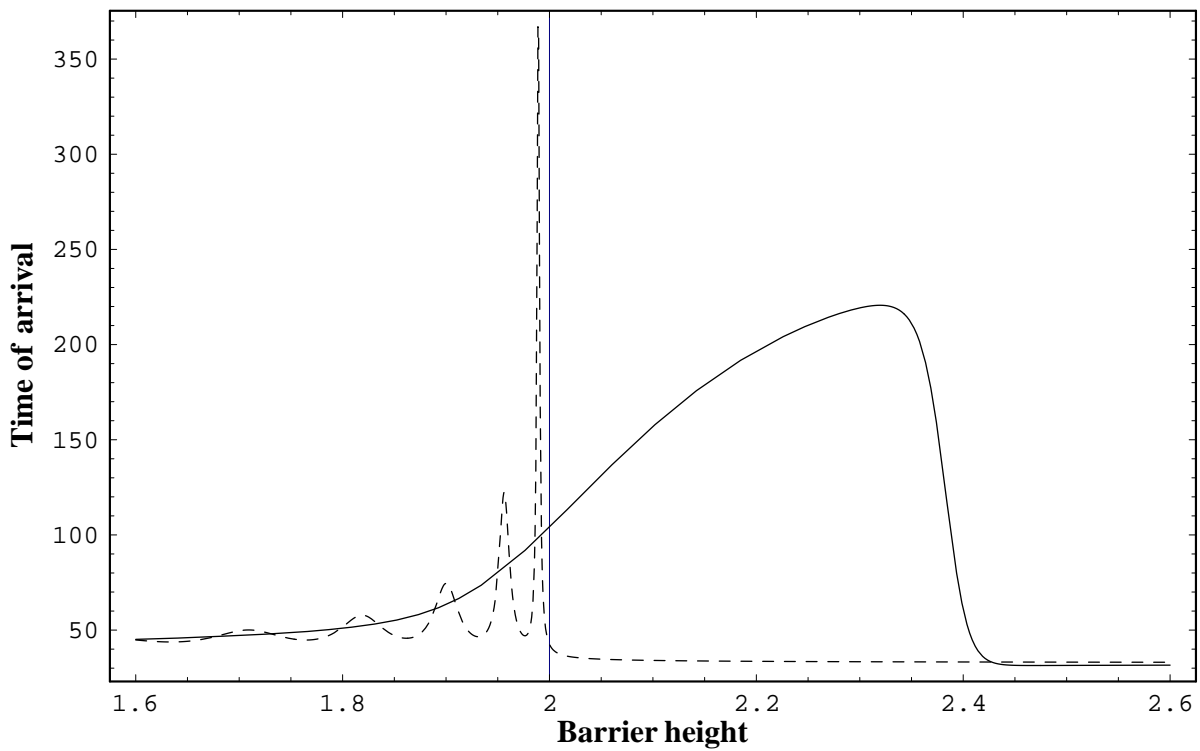}
 \end{center}
 \caption{Average time of arrival at the other side of a barrier
 of fixed width $a=15$  as a function of the barrier momentum $p_V$.
The parameters of
 the initial Gaussian  wave packet
are $q_0=-30,p_0=2,\delta=10,m=1$, and  the arrival at $x=50$,
 in  units with
 $\hbar=1$. The solid line is the quantum average (\ref{psi55}),
 the dashed line is the  phase time with
 momentum $p_0$. The asymptote to the left ($p_V\rightarrow 0$)
 is the time of arrival for free particles ($t_0=40$), the one to
 the right is  the Hartman time
$t_H(a)=t_0-\frac{m\, a}{p_0}=32.5$.
 } \label{fig1}
\end{figure}\nopagebreak

 We show the time of arrival at the other side of a barrier of momentum
 $p_V$ in the range $a=(1.6,2.6)$ in Fig. 1. For incident  plane waves
 with momentum $p_0$, the barrier would be crossed
 over for $p_V<p_0$, and tunneled through for $p_V>p_0$. Some
 retardation would be expected in the first case, just because the
 travel over the barrier would be slowlier than the free travel.
  This is clearly seen
 at the left of $p_0$ in the figure. Classically, the delay would
grow from zero (time $t_0$) to infinity as $p_V$ grows from $0$ to
$p_0$. The quantum behaviour is similar, with the oscillations of
the phase time swept away by the average that remains finite. To
the right of $p_0$, there is a dramatic difference between the
Wigner result, that inmediately sticks to the Hartman prediction
$t_H$, and the wave packet result, for which the time continues to
increase up to a certain barrier height and then, suddenly, drops
to $t_H$. This strange behaviour can be explained in the following
manner: Not being monoenergetic, the wave packet has momentum
components above and below $p_V$. The first of these cross above
the barrier, get retarded, and are responsible for the high time
value for $p_V$ just to the right of $p_0$. However, as the barrier
continues to grow, they become an ever lesser part of the packet.
The other parts of the packet (the components with momentum
$p<p_V$) tunnel through the barrier, and experience the Hartman
advance. They would arrive at $x$ in a time $t_H$. Their relative
importance in the wave packet increases steadily as $p_V$ continues
to grow and, eventually, they overcome the retarded components and
the process becomes pure tunneling. Then, the time of arrival drops
to $t_H$. We have numerically checked this behaviour, that we have
analyzed for several values of the barrier width in the range
(2,30). All the results are similar: Monotonic grow of the time
from $p_V=0$ (where $t=t_0$), up to $p_V\approx 2.5$, where $t$
drops suddenly to $t_H$. The general trend is a slow increase in
the value of the barrier momentum $p_V$ at which the drop takes
place,
 that shifts from about 2.2 to 2.7 as $a$ changes from 10 to 30. The
 maximum value of the time of arrival $\langle t_x \rangle$
 that is obtained just before
 the drop also increases; it is around 95 for $a=10$ and around 450 for
 $a=20$.

\begin{figure} \epsfxsize=16cm
 \begin{center}
 \ \epsffile{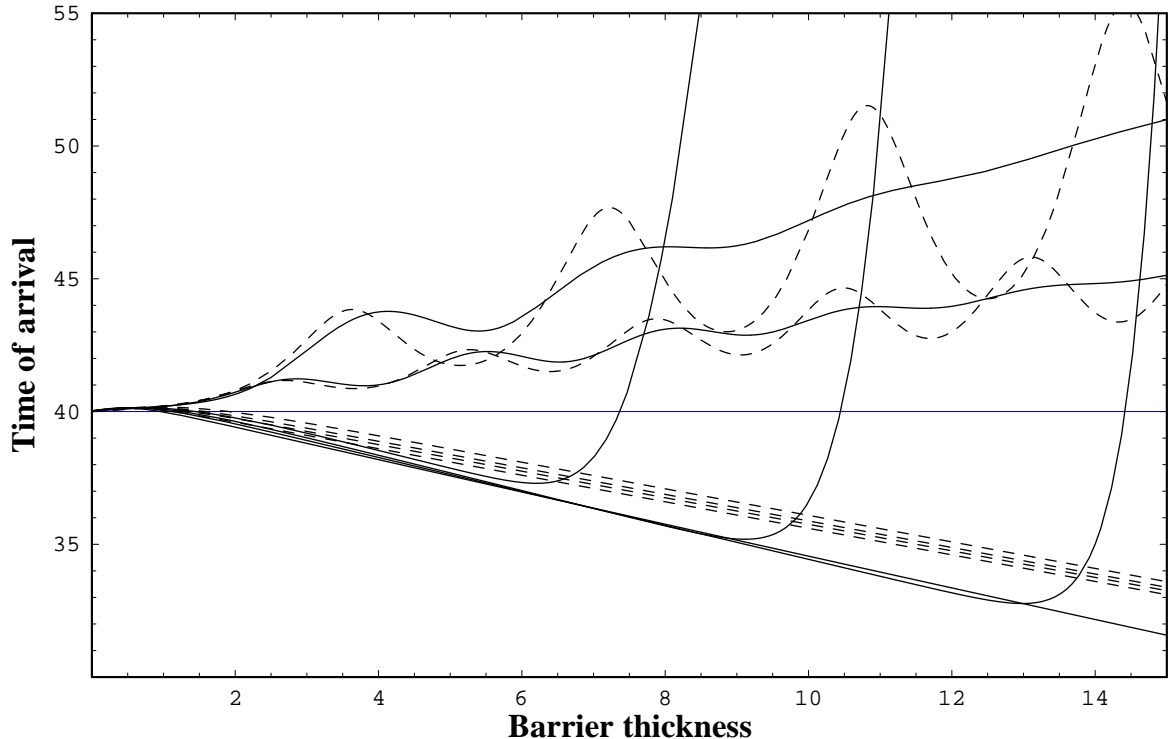}
 \end{center}
 \caption{Average times of arrival at the other side of a barrier
 as a function of the barrier width $a$. The initial wave packet is the
 same as in Fig.1, and again $x=50$. We show the predictions for
 $p_V=0,1.6,1.8,2.2,2.3,2.4,2.6$.
 The solid lines are the quantum averages (\ref{psi55}),
 the dashed lines are the corresponding Wigner times
 for a particle with momentum $p_0$ } \label{fig2}
\end{figure}\nopagebreak

 We show in Fig. 2 the average time of arrival and the Wigner (phase) time
as a function of the barrier width $a$ in the range $a=(0,15)$. We
display the predictions for different barrier heights
$p_V=0,1.6,1.8,2.2,2.3,2.4$ and $2.6$. For the free case
 ($p_V=0$, or $a=0$) all the results converge to $t_0=40$. We now
 disscuss the solid lines $\langle t_x \rangle$. The oscillatory
 curves above $t_0$ correspond to $p_V<p_0$. The get steeper as
 their momenta approach $p_0$ from below. The curves that stand
 partially below $t_0$ correspond to $p_V>p_0$ (tunneling). They
 share a similar behaviour: As the
  barrier width grows from $a=0$
the time of arrival decreases, practically saturating the Hartman
time $t(a)\approx t_H(a)=t_0-\frac{m\, a}{p_0}$. Then suddenly, at
a certain width (that increases with $p_V$), the average time jumps
dramatically to values that correspond to a long retardation. Note
that the jump for $p_V=2.6$ lies outside the range of the figure.
This behaviour is complementary to that shown in Fig. 1. Here, for
$p_V>p_0$ and moderate $a$, tunneling is the dominant phenomenon
and the time average tends to reproduce $t_H$. However, as the
barrier gets wider, tunneling gets more and more depressed. In
comparison, the intensity of the retarded components that pass over
the barrier is basically independent of $a$. They get relatively
more and more important and, eventually overcome tunneling, giving
rise to the observed transition. In practice, for wide enough
barriers, the probability of tunneling vanishes, and the other side
can be reached only by the very improbable and  very slow travel
over the barrier. This behaviour has been noticed independently
in~\cite{japanese}, and explained in the same way. In addition, we
have the tools to check these explanations. In particular, the
first product of our formalism is $P(t,x)$, the probability
distribution in times of arrival at $x$. Our numerical analysis for
$x=50$ and the different $p_V$'s and $a$'s that we are discussing
here show similar almost Gaussian shapes
 for these distributions, as correspond to the initial wave
packets chosen,  and similar widths for these  $P(t,x)$, whose
maxima are placed close to the corresponding mean values $\langle
t_x \rangle$. As expected, the  probabilities get numerically
smaller as the corresponding events  become more and more unlikely.
In short, these distributions give the best support for the
validity of the explanation offered here for this striking
behaviour, that can be understood only after weighting the obtained
time of arrival with the relative probability of the actual event
to which it corresponds.

\section{Quantum reflections}
Having analyzed the modifications introduced by the transmission
phenomena in the time of arrival at the other side of potential
barriers, we turn to the case of reflection. We divide the analysis
into the two seemingly different cases in which there is classical
reflection, and in which it is absent. The first case is
characterized by the presence of at least one turning point in the
path of the particle. The second one, by the absence of any of
them. Quantum mechanically there could be some transmission in the
first case, and some reflection in the second one. Accordingly, we
separate the disscussion that follows into the two  main disjoint
cases that cover all the possibilities. These are the case where
the potential energy grows to infinity somewhere (total
reflection), and the case where it is bounded everywhere (with
partial reflection and transmission).
\subsection{The case of total reflection}
The potential energy could grow unbound, thus reflecting any
conceivable incoming state. We consider here a monotonic potential
energy that vanishes for $q \to -\infty$ and goes to infinity for
$q\to \infty$ so that $\lim_{q\to +\infty} \langle q|E\rangle=0$.
This removes the degeneracy of the energy eigenstates. As no state
may arrive from the right,  $\langle q|E l (+)\rangle=0$. The
eigenstates $|E \rangle$ will contain the same amount of positive
and negative momenta, so that their asymptotic form normalized to
one traveling particle per unit time is $\lim_{q\to -\infty}
\langle q|E\rangle=\frac{1}{\sqrt{2\pi}}
(\frac{m}{2E})^{1/4}\cos(pq+\delta(E))$, where $\delta(E)$ is the
phase shift. This also fixes completely the eigenstates for finite
values of $q$.

The time of arrival at an arbitrary point $x$ is now
\begin{eqnarray}
\langle\psi|{\bf t}_x|\psi\rangle=&\frac{2\pi}{P(x)} \int dE\;
\sqrt{\frac{2E}{m}}\;|\langle x|E\rangle\;\langle
E|\psi\rangle|^2\;\times \nonumber\\ &\frac{d}{dE}\{ \arg\langle
x|E\rangle +\arg\langle E|\psi\rangle\}\label{n14}
\end{eqnarray}
which is the average of a quantity independent of $x$! This comes
about because in the present situation  the reflection coefficient
$R=\exp (-2 i\delta)$ is unimodular. Then, the net current density
vanishes, so that $\arg\langle x|E\rangle$ is independent of $x$.
This is the quantum version of the classical result that the sum of
the times of arrival at $x$ of the incoming and returning particles
is twice the {\bf toa} at the turning point and so, independent of
$x$. Obviously this ceases when $|R|$ becomes smaller than 1 (so
that the net current density is finite), something that is possible
only when $V$ is finite everywhere. Even then, the  classical
result is recovered from the quantum case in the limit $(E/V)<<1$
where $|R|\to 1$.

The individual times of arrival of the incoming and the returning
particles can be obtained straightforwardly by writing the enegy
eigenstates  as
\begin{equation}
\langle q |E\rangle=\frac{1}{\sqrt{2\pi}}(\frac{m}{2E})^{1/4}\;
M(q,E) \;\cos\phi(q,E)\label{n15}
\end{equation}
where $M$ is a real function with  $\lim_{q\to -\infty}M(q,E)=1$,
that vanishes faster than an exponential for $q\to +\infty$ to
satisfy the asymptotic form of the Schr\"odinger equation. The
state is thus written as the superposition at each point of an
incoming and a reflected wave with equal amplitudes, so that the
net current vanishes everywhere. The
 phase $\phi$ is
fixed by $\lim_{q\to -\infty}\phi (q,E)=pq+\delta(E)$ to match the
asymptotic form of the eigenstate disscussed above. Its derivative
gives the two opposite velocity fields $v_{\pm}(q,E)=\pm
\frac{d\phi(q,E)}{m\,dq}$ interfering at $q$. We recall that this
exact expression is valid for all the potentials of the form we are
considering here. The probability of ever arriving at $x$ and the
{\bf toa} can be given by straightforward application of
(\ref{k23}) and (\ref{k24}) by
\begin{eqnarray}
P(x)&=&\int dE\;M^2(x,E)  \cos^2\phi(x,E) |\langle
E|\psi\rangle|^2\label{n16}\\
 \langle \psi|{\bf t}_x|\psi\rangle
&=&\frac{1}{2P(x)} \int dE\;M^2(x,E)  \cos^2\phi(x,E)
\;\times\nonumber\\ &&|\langle E|\psi\rangle|^2\;
\left[t_{i}(x,E)+t_{r}(x,E)\right] \label{n17}
\end{eqnarray}
which is the weighted average over energies of the times of arrival
of the incoming and the reflected waves:
\begin{eqnarray}
t_{i}(x,E)&=&\frac{d}{dE}\{ \phi(x,E)+\arg\langle E|\psi\rangle\}
\label{n18}\\
 t_{r}(x,E)&=&\frac{d}{dE}\{ -\phi(x,E)+\arg\langle
E|\psi\rangle\}\label{n18a}
\end{eqnarray}
whose sum is explicitly $x$ independent.

To illustrate these results we consider now the case of a potential
that vanishes at the left of the origin and is linear at the right,
i.e. $V(q)= \theta(q) f q$, where $f$ is the force exerted on the
particle. This could be a model for a (charged) particle in a
constant electric field, or in the gravity field of the Earth. In
this case one gets $M$ and $\phi$ in terms of the Airy function Ai
and its derivative\nopagebreak Ai'.

\begin{equation}
M(q,E)=\left\{ \begin{array}{rr} 1&\;\;\;\;\mbox{for}\,\,\, q\leq 0
\nonumber\\ \sqrt{\frac{\mbox{Ai}[z]^2+(\frac{k_f}{p})^2
\mbox{Ai'}[z]^2}{\mbox{Ai}[z_0]^2+(\frac{k_f}{p})^2
\mbox{Ai'}[z_0]^2}}& \;\;\;\;\mbox{for}\,\,\, q> 0\end{array}
\right. \label{n19}
\end{equation}
where $z=k_f q-p^2/k_f^2, z_0=-p^2/k_f^2$ with $k_f=(2 m f)^{1/3}$.
For the phase one has
\begin{equation}
\phi(q,E)=\left\{ \begin{array}{rr}\arctan(-\frac{k_f
\mbox{Ai'}[z_0]}{p\,\,\, \mbox{Ai}[z_0]}\,)&\mbox{\,\,\,for}\,\,\,
q\leq 0 \nonumber\\
 \arctan(-\frac{k_f \mbox{Ai'}[z]}{p\,\,\,
\mbox{Ai}[z]}\,) &\mbox{\,\,\,\,for}\,\,\, q> 0\end{array}\right.
\label{n20}
\end{equation}
so the phase shift is given simply by $\delta(E)=\phi(0,E)$.

\begin{figure} \epsfxsize=16cm
 \begin{center}
 \ \epsffile{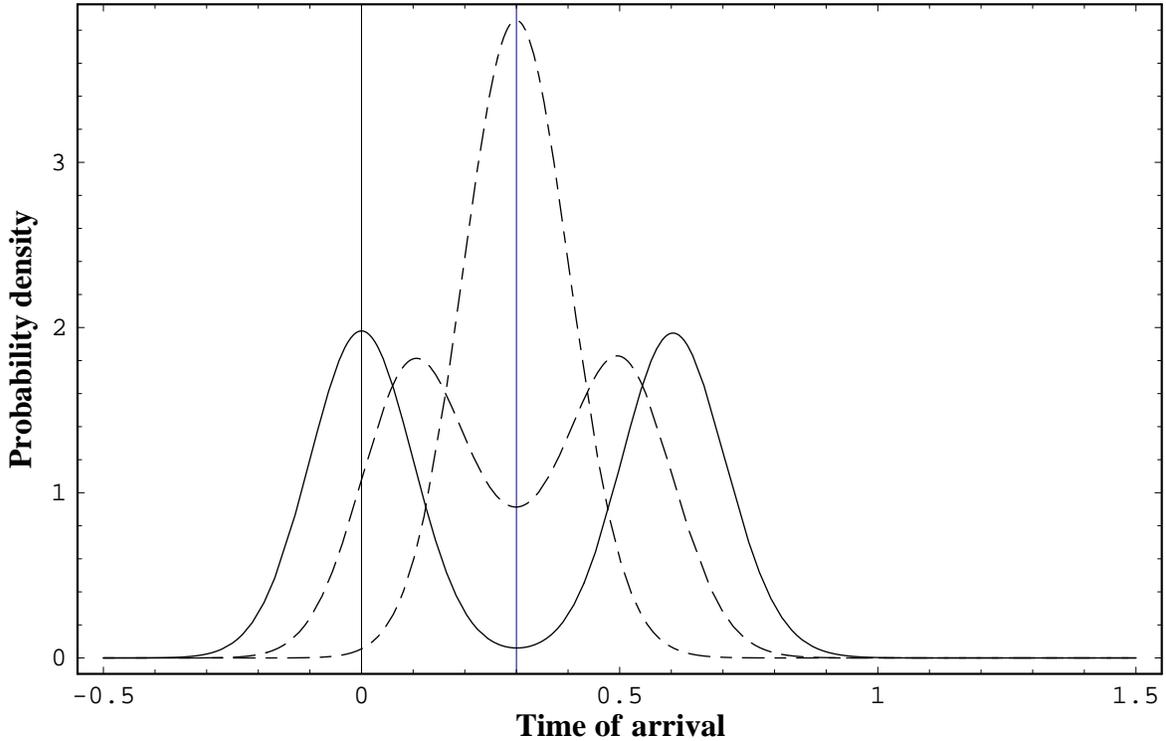}
 \end{center}
 \caption{ Probability distribution  $P(t,x)$ in times of arrival at
 $x$, for $x=q_0$ (solid line), $x=0.5\, q_0$ (dashed
 line) and at the classical turning point $x=E/f$ (dot-dashed line).
 The vertical lines correspond to $t=0$ and to the classical turning
 time respectively. The distributions are bimodal,
 with the two peaks corresponding to incidence and reflection
getting closer as $x$ approaches the turning point.} \label{fig3}
\end{figure}\nopagebreak

We present in Figures 3 and 4 our results for the the case of a
force of nominal value $f=100$, being the parameters of the initial
Gaussian wave packet (\ref{l8}) $q_0=-2,p_0=10,\delta=1$ and $m=1$.
For the normalized probability distributions in times of arrival
(\ref{k222}) we get pairs of peaks of equal heights - as correspond
to total reflection - that tend to merge into one as the detector
is displaced towards the classical turning point. This behaviour of
the peaks is also observed for the averaged times of arrival, that
follow the classical times. The small deviations from the parabolic
form  are negligible in comparison with the widths of the
distributions shown in Fig. 3. We have explored numerically the
details that change uninterestingly according to the values of
$f,p_0,\delta$ etc. so, we do not show them here. The general
picture is always the same: at the far left ($|q_0|>>E/f$) the
potential acts as an infinite height wall. The only sizeable
consequences of the actual strenght of the force are felt at
positions between the origin and the turning point, where they
resemble the classical effects. Part of this comes from the fact
that here position and energy combine into only a variable $q-E/f$.
But the resemblance arises because total reflection is always
present here, quantum as well as classically. This will be more
clear in the next section where we consider partial reflection that
lacks of classical analog.

\begin{figure} \epsfxsize=16cm
 \begin{center}
 \ \epsffile{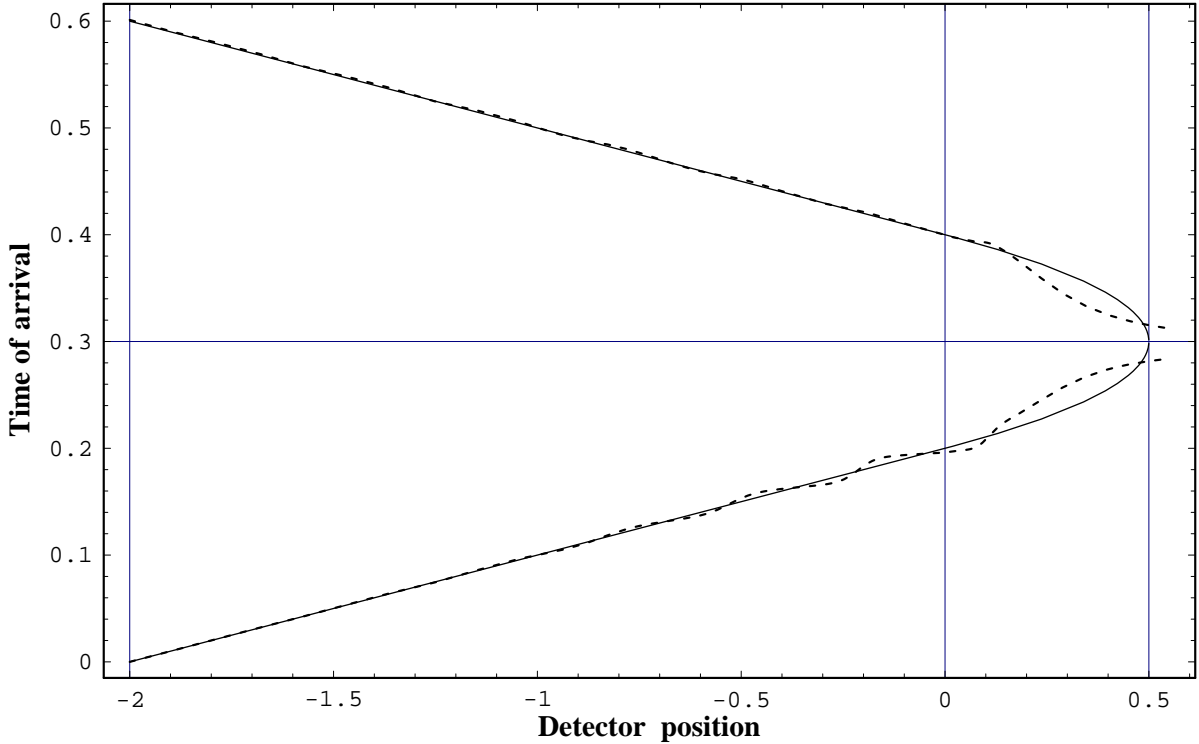}
 \end{center}
 \caption{ Average time of arrival for different detector positions.
 The vertical lines correspond to $x=q_0,\;\; x=0$ and $x=E/f$
 respectively.
 The solid line is the classical time, the dashed line is the
 quantum average of $t_i$ (lower part) and of $t_r$ (upper part).}
\label{fig4}
\end{figure}

\subsection{Partial reflections}
In classical mechanics  a potential interaction energy speeds up or
slows down the particles according to the local value of the force
$F(q)=-\frac{\partial V(q)}{\partial q}$. Accelerated or
decelerated, the particles continue to move along the same path
without reversing the direction. Only when one of them intercepts a
turning point (i.e. a point $q$ where $E=V(q)$) the particle
bounces back or, in other words, is reflected with probability
$P_R=1$. In the absence of these points, the particle is always
transmitted with probability $P_T=1$. Thus, most of the time
$P_T=1,P_R=0$. Only at the turning points $P_T=0,P_R=1$.

Quantum dynamics offers a very different perspective of the motion
of the particles. The Schr\"odinger equation implies that at every
point where the potential energy is finite, the particle is
partially transmitted and partially reflected, that is $0\leq
P_T\leq 1, 0\leq P_R\leq 1$, with $P_T+P_R=1$. The case of total
reflection analyzed in the previous section is one of close
correspondence between the classical and the quantum results, as we
shown there. Interesting departures from the classical behaviour
arise when there is no classical reflection. We will analyze this
case here.

To fix ideas, we consider a well behaved potential energy $V(q)\geq
0\; \forall$ finite $q$, that vanishes at the spatial infinity
faster than $q^{-1}$. In these conditions the energy eigenstates
can be written everywhere as a well defined superposition of
transmitted $\Phi_{tr}(q,E)$ and reflected $\Phi_{ref}(q,E)$ waves,
characterized by the positive or negative value of their currents:
$-\frac{i}{2 m}
(\Phi_{tr}^*\stackrel{\leftrightarrow}{\frac{d}{dq}}\Phi_{tr})\geq
0$, and  $-\frac{i}{2 m}
(\Phi_{ref}^*\stackrel{\leftrightarrow}{\frac{d}{dq}}\Phi_{ref})\leq
0$, with different amplitudes $|\Phi_{tr}|\neq|\Phi_{ref}|$ as
corresponds to this case of partial reflection. The eigenstates of
interest can be written as
\begin{equation}
 \langle q|E\, r\rangle=\sqrt{\frac{m}{2\pi p}} \{\Phi_{tr}(q,E)+
 \Phi_{ref}(q,E)\}\label{dic20}
 \end{equation}
 These waves are univocally determined by their asymptotic
 conditions, namely:
 \begin{eqnarray}
\lim_{q\to -\infty} \Phi_{tr}(q,E)=& e^{i p q},&  \lim_{q\to
+\infty} \Phi_{tr}(q,E)= T(E) e^{i p q}
 \nonumber \\
\lim_{q\to -\infty} \Phi_{ref}(q,E)=& R(E)
 e^{-i(p q+2 \delta(E))},&
\lim_{q\to +\infty} \Phi_{ref}(q,E)= 0\label{dic21}
 \end{eqnarray}
 as is the case for an incoming rightmover (\ref{dic20}). The
 results of the previous section are recovered in the limit where
 $T(E)\rightarrow 0$ which is the case only if the potential
 energy grows to infinity somewhere.

 If we prepare our initial Gaussian state $\psi(q)$ at a point
 $q=q_0$ where the potential energy is smooth enough, and keep the
 initial momentum $p_0>0$ large enough to consider
 $\tilde{\psi}(p)\approx 0$ for $p<0$, we can use the
 approximations
 \begin{equation}
 \langle E\, s|\psi\rangle \approx \delta_{rs} \sqrt{\frac{m}{p}}
\; \Phi_{tr}^*(q_0,E) \;|\tilde{\psi}(p)|\approx \delta_{rs}
\sqrt{\frac{m}{p}} \; e^{-ipq_0} \;|\tilde{\psi}(p)| \label{dic22}
 \end{equation}
 We have used the second of these already in Eq.(\ref{n1}). It is
 valid when $V(q)\approx 0$ for $q$ in the  $q_0$ neighbourhood where
 $\psi(q)$ is sizeable. We assume this is the case in what follows.

One of the deepest consequences of the superposition of transmitted
and reflected components that makes up the eigenstate
({\ref{dic20}) is that it leads to the inescapable presence of
interferences. In fact, the probability of presence at a point $q$,
and other quantities depending on it, contain the sum $|\langle q|E
r\rangle|^2\propto|\Phi_{tr}|^2+|\Phi_{ref}|^2+ 2 \Re(\Phi_{tr}
\Phi_{ref}^*)$, whose last term is the interference term. One could
say that, everywhere in its motion through the interaction region,
the quantum particle will be found in an evolving entangled state
of transmitted and reflected components. This can be traced back
mathematically to the continuity of the solutions of the
Schr\"odinger equation and of their first derivatives, and to the
associated Wronskian theorem. Physically, this may introduce all
sorts of interpretative difficulties in the analysis of particle
motion.

Summarizing, interferences pervade the realm of quantum motion.
They will show up in almost every quantum mechanical situation. Our
analysis of the time of arrival is not an exception. We have
avoided refering to them till now by focusing on very specific
cases. These were: The choice in Sect. 3.1. of a very smooth
potential analyzable semiclassically by the WKB method, that
neglects reflection. The analysis in Sect. 3.3. of the time of
arrival at points located at the other side of the barrier, where
$\Phi_{ref}=0$ so that any interference with the transmitted wave
vanishes. Finally, the analysis made in the previous section, where
we just ignored the effects due to the overlap of incoming and
reflected waves in $P(t,x)$, and the lack of a clear cut separation
between $t_i$ and $t_r$ in the presence of interferences. To be
precise, we dealt with reflection without paying the due attention
to these subtleties. We repair the ommission here.

The amplitude in time of arrival at a position $x$ within the
interaction range can be given by using (\ref{dic20}) and
(\ref{dic22}) in (\ref{psi2})
\begin{eqnarray}
\langle t x s|\psi\rangle &= &
\{A_{tr}(t,x)+A_{ref}(t,x)\}\nonumber\\ & \approx &
\frac{\delta_{sr}}{\sqrt{2\pi}} \int_0^\infty dp\sqrt{\frac{p}{m}}
\;|\tilde{\psi}(p)| \; e^{-i (Et+pq_0)}\;
\{\Phi_{tr}(x,E)+\Phi_{ref}(x,E)\}  \label{dic23}
\end{eqnarray}
This gives for the probability of ever arriving at $x$ Eq.
(\ref{k23}) the sum of three terms: The two separated probabilities
$P_{tr},P_{ref}$ of arriving with positive or with negative current
density, and a quantum interference term, whose presence deprives
the previous two of direct physical meaning. We thus get
$P(x)=P_{tr}(x)+P_{ref}(x)+I(x)$ with
\begin{equation}
P_{_{tr \atop ref}}(x)=\int dt\; |A_{_{tr \atop ref}}
 (t,x)|^2 \approx\int dp\;|\tilde{\psi}(p)|^2
 |\Phi_{_{tr \atop ref}}(x,E)|^2\label{dic24}
\end{equation}
and an interference term
\begin{eqnarray}
I(x)&=& 2\int dt \; \Re\{A_{tr}(t,x)\,A^*_{ref}(t,x)\}\nonumber\\
&\approx& 2 \int dp\; |\tilde{\psi}(p)|^2 \Re\{e^{-ipq_0}\;
\Phi_{tr}(x,E)\; \Phi_{ref}(x,E)^*\} \label{dic25}
\end{eqnarray}

The above quantities depend on the probabilities of transmission or
reflection from the initial position $q_0$ to the actual value $x$.
Consider a bounded potential barrier of finite range, but otherwise
arbitrary. Behind the barrier $P_{ref}$ vanishes, while $P_{tr}$ is
given by (\ref{psi44}) with a value independent of $x$, but
strongly dependent of $p_0,\delta$ and of the barrier's height and
width. For $x$ at the left of the barrier $\Phi_{tr}=e^{i p x}$
(what we are denoting as transmission is here incidence), but
$\Phi_{ref}=R(E)\, e^{-i( p x+2 \delta(E))},$ and only when there
is no reflection (no barrier) the intereferences dissappear. For
the total reflection case of the previous section, we get
$P_{tr}=P_{ref}$, while the interference term gives rise to the
term $\cos\{2(px+\delta(E))\}$ that builds up the factor
$\cos^2\phi$ that appears in (\ref{n16}) and (\ref{n17}). However
it does not prevent the definition of the quantities (\ref{n18})
and(\ref{n18a}) that allowed to split the {\bf toa} (\ref{n17})
into two positive contributions interpretable as the independent
$\langle t_x \rangle$ of an incoming packet and a reflected one
(Fig. 4).
 \begin{figure} \epsfxsize=16cm
 \begin{center}
 \ \epsffile{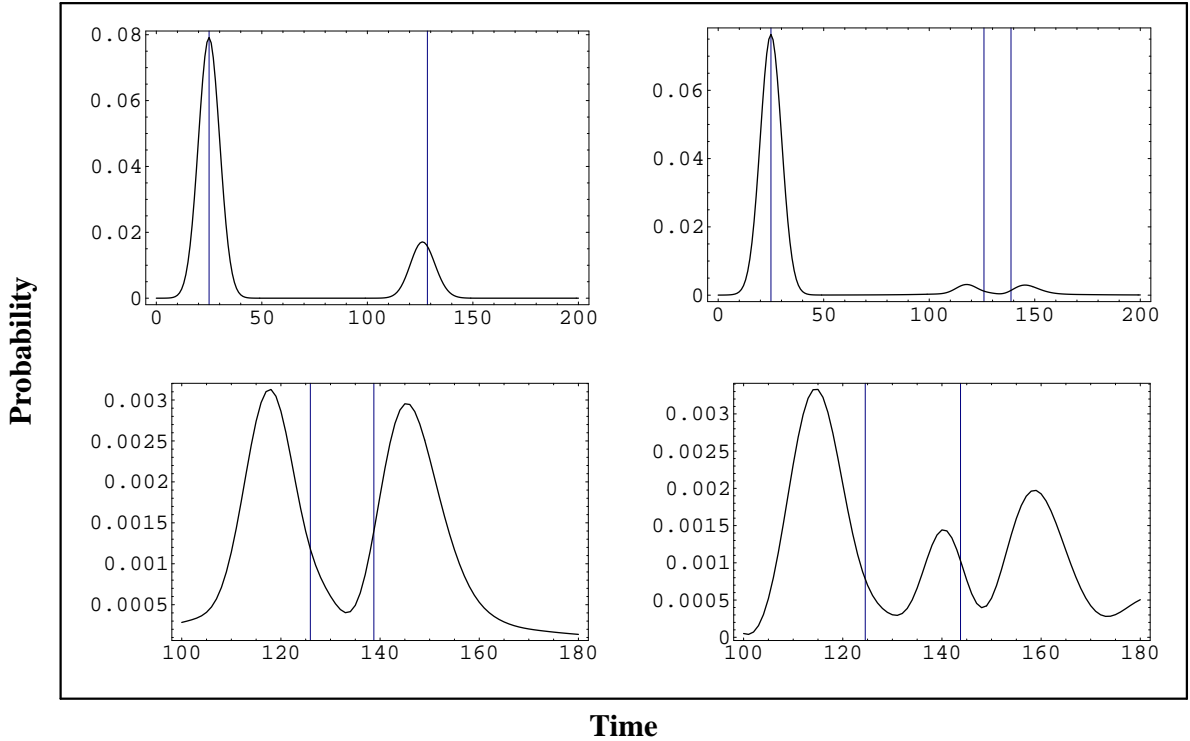}
 \end{center}
 \caption{Probability distribution $P(t,x)$ in times of arrival
 for reflection from finite potential barriers. The initial wave
 packet is the Gaussian one with $m=1,p_0=2,\delta=10$, placed at
 $q_0=-150$. The arrival position is at $x=-100$. The upper left
 figure is for a barrier width $a=4$ and $p_V=2.2$. At its right is
 the case $a=4$ and $p_V=1.9$, that is enlarged in the lower left
 part for the range $t=[100,180]$. An illustrative case of
 multimodal reflection distribution is shown at the lower right
 part, that corresponds to $p_V=1.9$ and $a=6$. The vertical
 grid lines correspond to the phase times of the incident wave
 $e^{i p_0 x}$ and of the reflected waves: $e^{-i p_0 x}$ for
 $p_V>p_0$, and the superposition $\sin p_0' a \; e^{-i p_0 x}$ for
 $p_V<p_0$.} \label{fig5}
\end{figure}

For finite barriers reflection is always present with an energy
dependent coefficient $R(E)<1$; it is less probable than incidence,
and tends to vanish as the barrier does. In Fig. 5 we give the
probability distributions of {\bf toa}  $P(t,x)$ at a point $x$,
whose bumps indicate, as in Fig. 3, the arrival of incident and
reflected parts of the time evolved initial wave packet. This is
the Gaussian one with $m=1,p_0=2,\delta=10$, placed at $q_0=-150$.
The arrival position is at $x=-100$, far from $q_0$ to avoid
interferences. The two upper figures are for a barrier of width
$a=4$. At the left is the case where $p_V=2.2$, and at the right
that with $p_V=1.9$. In both cases there is an incidence bump
centered at $t=m (x-q_0)/p_0=25$, and a structure to its right
corresponding to reflection. For $p_V=2.2$, and for all the cases
of total classical reflection ($p_V>p_0$), the latter is a
Gaussian-like bump shifted from the classical value at
$t=m(-x-q_0)/p_0=125$ by an amount $\langle
\frac{m}{p}(\frac{d\phi}{dp})\rangle$. However, for $p_V=1.9$ (in
general for $p_V<p_0$), the reflected distribution has a multi-bump
shape difficult to understand in terms of the phase time or of any
other approximation. In particular, neither the number of peaks,
nor their positions heights and widths can be approximated by
straight stationary phase methods. Two illustrative cases of these
shapes are shown in some detail in the two examples of the lower
part that correspond to $p_V=1.9$ and two close widths $a=4$ and
$a=6$.

\section{Conclusions}
We have worked out a formalism for obtaining the time of arrival at
a space point of particles that move through interacting media. Our
construction follows a circuitous path: we desist from first
computing the classical {\bf toa} of the problem, and then
quantizing it, a procedure that leads to a dead end. Instead, we
start from the quantum {\bf toa} of the free moving particle, and
then transform it canonically to the interacting case. This is
achieved by the use of the appropriate M\"oller operator that
implements the quantum version of the Jacobi-Lie canonical
transformation to free translation coordinates in phase space. In
the classical case we have the transformation of Eq.
(\ref{classic2}) whose quantum counterpart is
\begin{equation}
\{{\cal H}, H\}\stackrel{\Omega^\dagger}{\longrightarrow}\{{\cal
H}_0, H_0\}
 \label{conclussions1}
\end{equation}
where ${\cal H}_0$ and ${\cal H}$ are the Hilbert spaces of the
free and interacting particles, and $H_0, H$ the respective
Hamiltonian operators in these spaces. For simplicity, we have only
addressed explicitly cases in which the transformations are
unitary, which is the case when $\sigma(H)=\sigma(H_0)$. More
general situations that require of isometric transformations,
deserve a separate treatment by their physical relevance.

What we obtained here is a quantum formalism for the {\bf toa} in
terms of a POVM given by
\begin{equation}
P(t_1,t_2;x)=\sum_{s=r,l} \int_{t_1}^{t_2}dt\;|t\ x\
s\rangle\langle t\ x\ s| \label{conclussions2}
\end{equation}
which measures the probability of arrival at $x$ during the time
interval $(t_1,t_2)$. The normalized probability distribution
$P(t,x)$ was given in the Eq. (\ref{k222}) of Sect. II.B. Our
results are thus within the standard formalism of quantum mechanics
and can be interpreted in the standard way. There is nothing
special that singles out our theoretical predictions as unsuitable
for comparison with the experimental results. On the contrary, our
formalism predicts the result of actual experiments in the form of
numeric values and statistics for the recorded events.

After the definition and theoretical analysis of Sect. II. we have
performed explicit and complete calculations for the cases of an
unbounded linear potential, of the step potential and of the square
barrier. Our analysis of the quasi-classical case shows that in
this limit the {\bf toa} is simply given by the average of the
classical time of Eq. (\ref{j5}) over the quasi-classical wave
function. In the case of reflection, and for the arrival point
placed between the initial position of the wave packet and the
turning point ($x<0$), the probability distribution $P(t,x)$ is
governed by the quantum superposition of the incident $(A_{tr})$
and the reflected $(A_{ref})$ wave packets. In the case of total
reflection, where both are equally probable $P_{tr}(x)=P_{ref}(x)$,
we have obtained separate positive $\langle t_x \rangle$ even when
both amplitudes overlap. These were interpreted as the {\bf toa}'s
of the incident and reflected particles, and compared successfully
with the classical prediction. For partial reflection,
$P_{ref}(x)<P_{tr}(x)$ non overlapping amplitudes are necessary to
 to get separate average values for these times.
This problem is shared with the position and other operators. It is
not a defect of the formalism, but an effect of the interferences.
Fortunately enough, our formalism provides us with the probability
distribution $P(t,x)$ whose diverse humpy-bumpy shapes (Figs. 3 and
5) give the most complete information of the posible experimental
outcomes.

In the course of our numerical analysis we have detected that the
phase time $\tau_\phi$ not always gives a good approximation to the
most probable time of arrival. It provides a first estimate of the
time spent in the transmission or reflection, after substracting
the time of free flight. For transmitted wave packets we have
reobtained the advancement (i.e. a decrease in the {\bf toa}) in
the case of pure quantum tunneling. This  phenomenon, predicted by
Hartman long time ago~\cite{Hartman}, has been experimentally
evinced by the two photon experiments at
Berkeley~\cite{Steinberg3,Steinberg4} and the tunneling of optical
pulses at Wien~\cite{Spielmann}. However, our formalism predicts a
striking departure from the Hartman bound that we explain in detail
in Sect. IV. Our results for square barriers neatly show the
expected advancement roughly proportional to the width $\Delta t=-
m a/p$ (Figs. 1 and 2). However, whatever the mean energy $(E<V)$
of the incident wave packet, there is always a width $a_0$ such
that for $a>a_0$ the (very retarded) components of the packet that
stand above the barrier dominate over the (probabilistically very
depressed) tunneled ones, giving an overall effective strong
retardation. In other words, when the barrier is wide enough, its
width dominates over the Hartman lenght $\Delta x$ disscussed above
Eq. (\ref{n11}), that has a purely quantum origin. This restores
the classical expectation of no tunneling and very long delays.

We have also found other unanticipated phenomenon for purely
quantum reflection: the multiple bump structure that appears when
$p_V<p_0$. We have shown  in Fig. 5 this structure, that in some
sense is a counterpart of the interference pattern that appears in
multiple reflection of stationary waves. We think that this
feature, even if less spectacular than the superluminal tunneling
of photons, deserves experimental confirmation. An appropriate
modification of the two photon experiments could serve for this
purpose. It would require to place a quantum mirror in the path of
one of the entangled photons, and check for the presence (or
absence) of the multiple dip structure in the number of coincidence
counts predicted by the formalism.

All the examples above show that our construction of a quantum {\sl
toa} operator suitable for the presence of interactions allows the
exploration of many physical details in relevant situations. Its
extension to higher dimensional cases poses no conceptual
difficulties and opens the possibility of treating new questions.
Of great theoretical and experimental interest will be the
extension of this formalism to the cases in which the Hamiltonian
has bound states, where isometric (instead of simply unitary)
transformations will be requiered.

\section*{Appendix A}
 In the modern literature~\cite{mccauley}, a
classical Hamiltonian system with $n$ degrees of freedom is called
completely integrable ( a l\`a liouville) when it satisfies the
conditions $a$ and $b$ below:
\begin{itemize}
\item[a.] There are $n$ compatible conservation laws
$\Phi_i(q_1,\ldots,q_n,p_1\ldots,p_n;t)=C_i$, $i=1,\ldots,n$,
that is:
\begin{itemize}
\item[a.1.]$\dot{\Phi}_i
=\{\Phi_i,H\}+\frac{\partial\Phi_i}{\partial t}=0,\;\forall\;
i=1,\ldots,n.$
\item[a.2.] $\{\Phi_i,\Phi_j\}=0,\; \forall \;i,j=1,\ldots,n.$
\end{itemize}
\item[b.] The conservation laws define $n$ isolating integrals that
can be written as:
\begin{itemize}
\item[b.1.] $\Phi_i=C_i \Rightarrow
p_i=\phi_i(q_1,\ldots,q_n,C_1,\ldots,C_n;t),\; \forall\;
i=1,\ldots,n.$
\item[b.2.] $\frac{\partial \phi_i}{\partial q_j}=
\frac{\partial \phi_j}{\partial q_i}\, \;\forall\; i,j=1,\ldots,n.$
\end{itemize}
\end{itemize}
In these conditions, Hamilton equations define an integrable flow,
that is, a system of holonomic coordinates $(q(t),p(t))$ in phase
space for each instant of time:
\begin{eqnarray}
q_i(t)&=q_i(q_0,p_0;t),&\;\;i=1,\ldots,n. \label{classic1}\\
p_i(t)&=p_i(q_0,p_0;t),&\;\;i=1,\ldots,n.\nonumber
\end{eqnarray}
In other words, given a set of initial conditions $(q_0,p_0)$ of
the system, at each instant of time $t$ the system arrives at a
point $(q(t),p(t))$ in phase space. Conversely, these points define
the corresponding times of arrival. In this case, time meets the
requirements to qualify as a derived variable in phase space.

As Lie pointed out, for any arbitrary time there is a special
choice of coordinates in phase space  that mathematically
eliminates  the effects of interactions from these integrable
flows, (the new positions are ignorable coordinates). More simply,
integrable systems are canonically equivalent to a set of
translations (or circular motions) at constant speed. It is
customary to denote the variables that determine these translations
as action-angle variables, which strictly is appropriate only in
the case of periodic systems, where the (closed) flow lines are
topologically equivalent to circles.

For integrable flows, there is a canonical transformation (the
Jacobi-Lie transformation)
\begin{equation}
\{q,p;H(q,p)\}\stackrel{W(q,P)}{\longrightarrow}\{Q,P;\bar{H}
(Q,P)\}
 \label{classic2}
\end{equation}
with $H(q,p)=\bar{H}(Q,P)$, that gives the free translation
coordinates $P(t)=P$, and $Q(t)=\frac{P}{m} t +Q$ of the
translation flow with $\bar{H}(Q,P)=\frac{P^2}{2m}$, in terms of
the coordinates and momenta $(q(t),p(t))$ of the actual flow with
$H(q,p)=\frac{p^2}{2 m}+V(q)$. This transformation is of the form
$W(q,P)$, that is, a function of the old coordinates and the new
momenta, so that
\begin{equation}
Q=\frac{\partial W}{\partial P},\;\; p=\frac{\partial W}{\partial
q} \label{classic3}
\end{equation}
Finally, $W$ can be obtained explicitly as a complete integral of  the
Hamilton-Jacobi equation:
\begin{equation}
H(q,\frac{\partial W}{\partial q})=\frac{P^2}{2 m}\label{classic4}
\end{equation}
Now, the canonical relation among the new and the old variables is:
\begin{eqnarray}
P&=&\mbox{sign}(p)\sqrt{2 m\; H(q,p)}\nonumber\\
 Q&=&\int_0^{q} \frac{dq'}{\sqrt{1-\frac{V(q')}{H(q,p)}}}+Q_0\nonumber
 \end{eqnarray}
where $Q_0$ is a constant. As a byproduct, time gets defined in
equivalent manner in terms of the old variables, or of the new
ones. If the particle arrives at $q(t)=x$ in the instant $t(x)=t$,
then:
\begin{equation}
t(x)=\frac{m}{P}(X-Q)=\mbox{sign}(p)\int_q^x\frac{m \;dq'}{\sqrt{2
m\,(H(q,p)- V(q'))}} \label{classic5}
\end{equation}
where $X=\partial W(x,P)/\partial P$ (obviously, $X=Q(t)$ by
construction). This duality, devoid of practical interest in the
classical domain, is at the foundations of the quantum method
developed in this paper. Finally, note that for simplicity we have
specialized the notation to the case of autonomous Hamiltonian
systems with only one degree of freedom, all of them trivially
integrable ($H(q,p)=E$ being the needed conserved quantity).


\section*{Appendix B}

For free particles Eq. (\ref{j5}) gives $t_{x 0}(q,p)=m (x-q)/p$
that, in spite of its simplicity, presents some problems for
quantization~\cite{Aharonov1,Tate,Juan} whose solution we outline
here. First of all, it requires symmetrization:
\begin{equation}
{\bf  t}_{x 0}({\bf q},{\bf p})=m (\frac{x}{{\bf p}}- \frac{1}{2}
\{{\bf q},\frac{1}{{\bf p}}\}_+)=-e^{-i{\bf p}x}\sqrt{\frac{m}
{{\bf p}}}{\bf q}\sqrt{\frac{m}{{\bf p}}}e^{i{\bf p}x} \label{jj}
\end{equation}
As is well known, the eigenstates $|t x s 0\rangle$ of this
operator in the momentum representation can be given as ($\hbar
=1$)
\begin{equation}
\langle p|t x s 0\rangle=\theta(sp) \sqrt{\frac{|p|}{m}}\,\exp(i
\frac{p^2}{2m}t)\,\langle p | x \rangle \label{k155}
\end{equation}
where we use $s=r$ for right-movers ($p>0$), and $s=l$ for left
movers ($p<0$.) The label $0$ stands for free case. Finally, the
argument $sp$ of the step function that appears in the momentum
representation is $+p$ for $s=r$, and $-p$ for $s=l$. The
degeneracy of the energy with respect to the sign of the moment is
explicitly shown by means of the label $s$ in the energy
representation, where
\begin{equation}
\langle E s' 0|t x s 0\rangle = \delta_{s's}\,(\frac{2 E}{m})^{1/4}
e^{i E t}\, \langle E s 0 |x\rangle \label{k15}
\end{equation}

Summarizing, there is a time (of arrival at $x$) representation
spanned by the eigenstates
\begin{equation}
|t x s 0\rangle=(\frac{2 H_0}{m})^{1/4} e^{i H_0 t} \Pi_{s0}
|x\rangle \label{k15a}
\end{equation}
where $\Pi_{s0}$ projects on the subspace of right-movers ($s=r$),
or left-movers ($s=l$), i.e.
\begin{equation}
\Pi_{s0} =\int_0^{\infty} dE |E s 0\rangle \langle Es 0|
\label{k15b}
\end{equation}

These time eigenstates are not orthogonal, which in the past gave
rise to serious doubts about their physical meaning. The origin of
this problem can be traced back to the fact that (\ref{jj}) is not
self-adjoint, that is $\langle\varphi|{\bf t}_{x 0} \psi\rangle
\neq \langle{\bf t}_{x 0} \varphi| \psi\rangle$.  This was proved
by Pauli~\cite{Pauli} long time ago and is due to the lower bound
on the energy spectrum. The problem emerges as soon as one attempts
integration by parts in the energy representation.
Ref.~\cite{Muga6} is a recent illuminating review of these and
other related questions.

The measurement problem posed by this not self-adjoint {\bf toa}
operator can be solved by interpreting it in terms of a Positive
Operator Valued Measure (POVM), that only requires the hermiticity
of ${\bf t}_{x 0}$ (i.e. ${\bf t}_{x 0}={({\bf t}_{x 0})^*}^\top$).
Here, instead of a Projector Valued spectral decomposition of the
identity operator, one has the POVM
\begin{eqnarray}
P_0(\Pi(x);t_1,t_2)&=& \sum_s \int_1^2 dt \, |t x s 0\rangle
\,\langle t x s 0|\label{j6}  \\ &=&
 \sum_s \int_1^2 dt \,(\frac{2 H_0}{m})^{1/4}\, e^{i H_0 t}\,
 \Pi_{s0}\,\Pi(x)\,\Pi_{s0}\, e^{-i H_0 t}\,
 (\frac{2 H_0}{m})^{1/4}\nonumber
\end{eqnarray}
where $\Pi(x)=|x\rangle\, \langle x |$ is the projector on $x$.
Here, $P_0(1,2)^2\neq P_0(1,2)$ because $|t x s 0\rangle \langle t
x s 0|$ is not a projector, as the states are not orthogonal, but
where the limit as $t\rightarrow \infty$ of $P_0(-t,+t)$ is the
identity. The attained time operator is no longer sharp, but is
well suited for measurement. This solution has been implemented
in~\cite{Giannitrapani1}, and extensively analyzed in refs.
~\cite{Giannitrapani2,Toller1} and in the review~\cite{Muga6}. In
this POVM formulation the {\bf toa} is given by the spectral
decomposition
\begin{equation}
{\bf t}_{ 0} (H_0, \Pi(x))=\int_{-\infty}^{+\infty} dt\,  t\,
(\frac{2 H_0}{m})^{1/4}\, e^{i H_0 t}\, {\cal P}_0(x)\, e^{-i H_0
t}\, (\frac{2 H_0}{m})^{1/4}\label{j6a}
 \end{equation}
 where ${\cal P}_0(x)=\sum_s \Pi_{s0}\, \Pi(x)\, \Pi_{s0}$,
 which is not a projector.


\end{document}